# Semi-Quantitative Analysis and Serological Evidence of Hepatitis A Virus IgG Antibody among children in Rumuewhor, Emuoha, Rivers State, Nigeria

**Okonko IO[1*], Adim CC[1], Okonko BJ[2], & Mba EI[3]**
[1]Virus & Genomics Research Unit, Department of Microbiology, University of Port Harcourt, Port Harcourt, Nigeria.
*Corresponding author's E-mail address: iheanyi.okonko@uniport.edu.ng; Tel: +2347069697309, ORCID iD: 0000-0002-3053-253X
[2]Virology & Immunology Research Unit, Department of Applied Microbiology, Ebonyi State University, Abakaliki, Nigeria
[3]Enviromental Microbiology & Bioinformatics Research Unit, Department of Microbiology, Madonna University, Elele, Rivers State, Nigeria.

**Abstract**
Hepatitis A virus (HAV) infection has been greatly reduced in most developed countries through the use of vaccine and improved hygienic conditions. However, the magnitude of the problem is underestimated and there are no well-established Hepatitis A virus prevention and control strategies in Nigeria. The aim of this study was to determine the prevalence of Hepatitis A virus infection among children aged 2 to 9 years in Rumuewhor, Emuoha LGA, Rivers State, Nigeria. Blood samples were collected from the 89 children enrolled in this study, and analyzed for the presence of HAV IgG antibodies using ELISA techniques. Of the 89 participants, 22 (24.7%) tested positive for HAV IgG antibodies, while 67 (75.3%) were negative. The children within the ages of 4 – 6 years had the highest seropositivity rate (33.3%) while those less than 4 years had the least seropositivity rate (22.4%). The prevalence rate ratio of the males to females was 1:1.3. There was no significant difference ($p>0.05$) between IgG seropositivity and age groups and gender. However, there was a statistical association of IgG seropositivity rates with respect to immunization. The seroprevalence rate recorded in this study was significant, indicating that the virus is endemic in this study area. Proper awareness, health education and vaccination are imperative to controlling and preventing HAV infection in Rumuewhor, Emuoha, Rivers State, Nigeria.

## 1. INTRODUCTION
Hepatitis A virus (HAV) is one of the enterically transmitted viruses, affecting both children and adults worldwide, especially in developing countries like Nigeria. It occurs sporadically with a trend of cyclic recurrences (Guzman-Holst et al., 2022) and contribute significantly to morbidity and mortality globally. HAV infection has been reported to be associated to contaminated water and food as well as poor personal hygiene and sanitation (Tahaei et al., 2012; Patterson et al., 2019). According to World Health Organization (WHO), hepatitis A caused approximately 7134 deaths in 2016, which accounts for 0.5% of the mortality due to viral hepatitis (WHO, 2019).
HAV infection clinically presents as fever, malaise, nausea, abdominal discomfort, dark urine, anorexia, and jaundice (Guzman-Holst et al., 2022). Previous reports have shown that children less than 6years of age do not show these symptoms and hence, develop a lifelong protective immunity (Aggarwal et al., 2015). However, older children and adults have been documented to have symptomatic illnesses which may eventually lead to an acute infection, hepatic damage and jaundice (Murphy et al., 2013; lemon et al., 2017).

Specific antibodies as a result of the body's immune response to HAV infection can be detected and used to measure the burden of the disease. Clinical routine diagnosis of hepatitis A is made by finding anti-HAV IgM and anti-HAV IgG in the serum of the patients while an option is the detection of the virus and/or antigen in the feces (Waje et al., 2017). Immunoglobulin M (IgM) anti-HAV antibody is used to measure acute/recent infection while immunoglobulin G (IgG) anti-HAV antibody measures past infection (Aggarwal et al., 2015).

According to the review by Jacobsen & Wiersma, 2010, Africa was classified as a high HAV endemic region. Serological surveys from different socio-economic backgrounds and population, showed HAV seropositivity rates varying from 45% to 90% (Almuneef et al., 2006; Fathalla et al., 2000; Ikobah et al., 2015; Okara et al., 2017; Ekanem et al., 2018). Though other countries in the developed world have experienced a decrease in HAV seroprevalence rate, Ekanem et al., 2018 claimed that there is no significant decrease in HAV infection in African children, in the past few decades.

Based on World Health Organization, the most effective way to prevent HAV infection is to improve sanitation and vaccination. Unfortunately, due to lack of proper awareness and education, that goal has not been reached. This study was designed to determine the IgG seroprevalence of Hepatitis A virus among children residing in Emuoha, Rivers State, Nigeria and to ascertain the associated demographic risk factors influencing the prevalence of HAV in the study area.

## 2. MATERIALS AND METHODS
### 2.1 Study Area
This investigation was conducted by enrolling patients attending Primary Health Care Centre (PHC) Rumuewhor, Emohua Local Government Area. Rumuewhor is a rural area made up of four villages. Its inhabitants are majorly farmers.

### 2.2 Study Design
This study was a cross sectional, hospital-based study which sought to detect and evaluate the immunoglobulin G antibody against Hepatitis A Virus among children residing attending Primary Health Care Centre, Emuoha, Rivers State.

### 2.3 Eligibility Criteria
A consent form including a structured questionnaire was given to consenting parents and guardians of children within 2 to 9 years, attending the Primary health Care Centre in Rumuewhor, Emuoha, Rivers State.

### 2.4 Study Population
The target population involved children aged 2 to 9years attending the Primary Health Care Centre Rumuewhor for a routine immunization and routine check-up within the period of study. A total sample size of 89 children were randomly selected and enrolled in this study.

### 2.5 Sample Collection
Three millimeters of blood sample was aseptically collected by venipuncture from each participant into EDTA bottles. The blood samples were centrifuged at 3000rmp for 5 minutes and the plasma was obtained and aspirated into new sterile Eppendorf tubes and properly labelled. The plasma samples were transported in cold chain to the laboratory and stored at $-20^0$C till analysis.

### 2.6 Serological Analysis of HAV IgG Antibody
Plasma samples were analyzed for HAV antibodies using the ELISA kit manufactured by Dia.Pro Diagnostic Bioprobes Srl, Milano, Italy. The entire analysis was performed according to the manufacturer's instructions. Results were calculated and interpreted. For HAV IgG antibody, samples with S/Co less than 0.9 were regarded as negative, while those with S/Co greater than 1.1 were positive.

### 2.7 Data Analysis
Data was analyzed using the Statistical Package for Social Science version 22.0 for windows. Chi-square test was used for analyzing the qualitative variables and to test for association where appropriate. A p value of <0.05 was considered significant.

### 3. RESULTS
### 3.1. Characteristics of the Participants
Eighty-nine children aged 2-9 years participated in this study. The socio-demographic data for the samples were stratified and shown in Table 1. The age group 2- 4 years constituted the largest populations making up 75.3% (n=67). This was followed by 4 – 6 years which constituted 16.9% (n=15) while age group 7-9 years was least, representing 7.8% (n=7) of the total population. Fifty (56.2%) of the samples were males and 39 (43.8%) were females, giving a male to female ratio of 1:1.3. In addition, children immunized among the enrolled participants were 20 (22.5%) while those not immunized were 69 (77.5%)

### 3.2 Overall seroprevalence of Hepatitis A virus IgG antibody
Twenty-two children tested positive for anti-HAV IgG antibody giving a prevalence rate of 24.7%. Thus, those that were seronegative were 67 (75.3%) as shown in Table 1.

### 3.3 Seroprevalence of HAV IgG antibody among children in relation to age group
Age group 4 - 6 years had the highest seropositivity rate of 33.3% while age group 7 – 9 years had 28.6%. The lowest seropositivity rate was obtained among children <4 years of age (Table 1). No significant difference was observed in relation to their age (P = 0.65).

### 3.4. Prevalence of HAV IgG antibody among children in relation to their gender
Males had HAV IgG seroprevalence of 22% which is lower than the 28.2% obtained among the females (Table 1). There was no significant difference between the seropositive rates and their gender (P = 0.50)

### 3.5. Prevalence of HAV antibodies among children in relation to immunization

Among those immunized previously, 45.0% were seropositive for HAV IgG antibody while 18.0% of those not immunized had anti-HAV IgG antibody in their plasma. IgG seropositivity was statistically associated with the immunization of the children ($p<0.05$) as shown in Table 1.

**Table 1: Prevalence of HAV IgG antibody in relation to the socio-demographic characteristics of the children aged 2 – 9 years**

| Socio-Demographic Characteristics | Category | No. Tested (%) | No. Positive (%) | P value |
|---|---|---|---|---|
| Age (years) | < 4 | 67 (75.3) | 15 (22.4) | 0.65 |
|  | 4 – 6 | 15 (16.9) | 5 (33.3) |  |
|  | 7 – 9 | 7 (7.8) | 2 (28.6) |  |
| Sex | Males | 50 (56.2) | 11 (22.0) | 0.50 |
|  | Females | 39 (43.8) | 11 (28.2) |  |
| Immunization status | Immunized | 20 (22.5) | 9 (45.0) | 0.02 |
|  | Unimmunized | 69 (77.5) | 13 (18.8) |  |
| **Total** |  | **89 (100.0)** | **22 (24.7)** |  |

### 4. DISCUSSION

The determination of the seroprevalence of HAV in samples has been useful in studies involving its transmission and persistence. This study showed that 24.7% of the children, in Rumuewhor, Emuoha, Rivers State, had previous contacts with hepatitis A virus leaving a larger proportion of the children seronegative (75.3%). This result obtained, agrees with the study of Ikobah et al., 2015 and Ekanem et al., (2018), demonstrating that areas of low socioeconomic conditions have an increased risk of HAV infection compared to those in the high socioeconomic groups.

The HAV IgG seroprevalence in this study was lower than those reported in previous studies conducted in Saudi Arabia, India, Cambodia and South Africa (El-Gilany et al., 2010; Gupta et al., 2019; Nagashima et al., 2021; Plessis et al. 2022). It was also lower than the result obtained in some parts of Nigeria (Ikobah et al., 2015; Ekanem et al., 2018). However, it was higher than the study carried out by Okara et al. (2017) in Abuja in which the seroprevalence rate obtained was 2.9%. Differences in the geographic area, study population, living standard, sanitary levels, environmental hygiene, underreporting, the reluctance of patients seeking medical care, sample types and other socioeconomic conditions (Ghorbani et al., 2010; Okara et al., 2017; Palewar et al., 2022) could be the reason for such wide variation in positivity rates.

The positive detection rate of IgG seroprevalence was highest in children aged 4 – 6 years (33.3%). This finding was similar to the prevalence rate obtained in the study conducted by EL-Gilany et al and lower than the result obtained by Gupta et al., (2019). Children below the age of 4years had the lowest seropositivity rate in this study, thus, agreeing with previous works conducted by Du Plessis et al. (2022) and Wang et al. (2001). Ikobah et al. (2015) claimed that cumulative effect over years could influence higher seropositivity among the older age group, hence, confirms the fact that anti-HAV IgG increases with increased age (Daftary & Patel, 2021).

Findings in this study revealed that anti-HAV IgG seroprevalence was significantly higher among the females when compared to their male counterparts. This corroborates with the study done in Mexico where the HAV seroprevalence was 51% in the females and 49% in the males (Guzman-Holst et al., 2022) and disagrees with the study conducted by Daftary & Patel, (2021) and Okara et al. (2017) in which the seroprevalence was higher in males. The prevalence rate ratio of the males to females is 1:1.3. HAV is transmitted through the faecal-oral route and spreads more in places and among people with poor hygienic practices. It is likely possible that the females were more exposed.

HAV infection risk has been associated with immunization in this study. The highest seropositivity rate (45%) was obtained among children that were immunized, indicating that they have been previously exposed to HAV infection or enjoy the protection conferred on them by other vaccines. Also, about 18% of the children not immunized were seropositive, demonstrating that they too, may have been exposed to several bouts of the viral infection. Thus, the higher the number of persons immunized, the lower the risk of infection. Though HAV infection in children is typically an acute illness associated with general and nonspecific symptoms (Quiros-Tejeira, 2022), its self-limiting nature may account for the presence of the IgG antibodies circulating in the plasma of the children enrolled in this study.

## 5. CONCLUSION
Findings from our evaluation indicated that many children were anti-HAV IgG seropositive. This could be as a result of the low socioeconomic status and the likely poor hygienic practices in the study area, since most of the inhabitants are farmers. Proper health education, adequate sanitation and vaccination can improve health conditions and result in a decline in the disease prevalence.

**Compliance with ethical standards**
*Acknowledgements*
The management and staff of the Primary Health Care Centre Rumuewhor in Emohua Local Government Area of Rivers State, Nigeria, Nigeria, assisted the authors during the enrollment process and sample collection for this study. The authors appreciate the subjects' willingness to participate in the study and Mrs. Mercy Elenwo for the and blood sample collections.

*Disclosure of conflict of interest*
The authors claim that there are no conflicting interests.

*Statement of ethical approval*
According to all authors, the University of Port Harcourt Research Ethics committee evaluated and approved all experiments. The investigation is therefore carried out following the moral principles outlined in the 1964 Declaration of Helsinki.

*Statement of informed consent*

"All authors state that the parents of all the children who participated in the study gave their informed, voluntary consent."

**REFERENCES**

Aggarwal, R., & Goel, A. (2015). Hepatitis A: epidemiology in resource-poor countries. *Current opinion in infectious diseases*, *28*(5), 488-496.

Almuneef, M. A., Memish, Z. A., Balkhy, H. H., Qahtani, M., Alotaibi, B., Hajeer, A., ... & Al Knawy, B. (2006). Epidemiologic shift in the prevalence of Hepatitis A virus in Saudi Arabia: a case for routine Hepatitis A vaccination. *Vaccine*, *24*(27-28), 5599-5603.

Daftary, N., & Patel, D. (2021). Prevalence of Hepatitis A Virus (HAV) and Hepatitis E Virus (HEV) In the Patients Presenting With Acute Viral Hepatitis at a Tertiary Care Hospital, Rajkot, Western India.

Du Plessis, N. M., Haeri Mazanderani, A., Motaze, N. V., Ngobese, M., & Avenant, T. (2022). Hepatitis A virus seroprevalence among children and adolescents in a high-burden HIV setting in urban South Africa. *Scientific Reports*, *12*(1), 20688.

Ekanem, E. E., Ikobah, J. M., & Okpara, H. C. (2018). Faeco-orally transmitted viral hepatitis in african children. *Journal of Immunological Sciences*, *2*(2).

El-Gilany, A. H., Hammad, S., Refaat, K., & Al-Enazi, R. (2010). Seroprevalence of hepatitis A antibodies among children in a Saudi community. *Asian Pacific Journal of Tropical Medicine*, *3*(4), 278-282.

Fathalla, S. E., Al-Jama, A. A., Al-Sheikh, I. H., & Islam, S. I. (2000). Seroprevalence of Hepatitis A virus markers in Eastern Saudi Arabia. *Saudi Medical Journal*; 21(10), 945–9

Ghorbani, G., Mahboobi, N., Lankarani, K. B., & Alavian, S. M. (2010). Hepatitis A Prevention Strategies, Haiti Case: Should Rescuers Be Immunized. *Iran Red Crescent Medical Journal*, 12: 221-223

Gupta, P., Chauhan, S., Agarwal, J., Jain, A., Sawlani, K. K., Goel, A. & Himanshu, D. (2019). Status of adult immunity to hepatitis A virus in healthcare workers from a tertiary care hospital in north India. *The Indian Journal of Medical Research*, *150*(5), 508.

Guzman-Holst, A., Luna-Casas, G., Burguete Garcia, A., Madrid-Marina, V., Cervantes-Apolinar, M. Y., Andani, A. & Sánchez-González, G. (2022). Burden of disease and associated complications of hepatitis a in children and adults in Mexico: A retrospective database study. *PLoS One*, *17*(5), e0268469.

Ikobah, J. M., Okpara, H. C., Ekanem, E. E., & Udo, J. J. (2015). Seroprevalence and predictors of hepatitis A infection in Nigerian children. *Pan African Medical Journal*, *20*(1).

Lemon, S. M., Ott, J. J., Van Damme, P., & Shouval, D. (2017). Type A viral hepatitis: A summary and update on the molecular virology, epidemiology, pathogenesis and prevention. *Journal of Hepatology.*

Nagashima, S., Ko, K., Yamamoto, C., Bunthen, E., Ouoba, S., Chuon, C., ... & Tanaka, J. (2021). Prevalence of total hepatitis A antibody among 5 to 7 years old children and their mothers in Cambodia. *Scientific Reports*, 11(1), 1-7.

Okara, G. C, Hassan S., & Obeagu, E. I. (2017). Hepatitis A virus infection among apparently healthy Nigerian Subjects. *Journal of Biomedical Science*, *6*, 2.

Palewar, M. S., Joshi, S., Choudhary, G., Das, R., Sadafale, A., & Karyakarte, R. (2022). Prevalence of Hepatitis A virus (HAV) and Hepatitis E virus (HEV) in patients presenting with acute viral hepatitis: A 3-year retrospective study at a tertiary care Hospital in Western India. *Journal of Family Medicine and Primary Care*, *11*(6), 2437-2441.

Patterson, J., Abdullahi, L., Hussey, G. D., Muloiwa, R., & Kagina, B. M. (2019). A systematic review of the epidemiology of hepatitis A in Africa. *BMC infectious diseases*, *19*(1), 1-15.

Quirós-Tejeira, R. E., Edwards, M. S., Rand, E. B., & Hoppin, A. G. (2016). Overview of hepatitis A virus infection in children. *At UpToDate: http://www. uptodate. com Last updated: Mar*, *31*, 2016.

Tahaei, S.M. E., Mohebbi, S. R., & Zali, M. R. (2012). Enteric hepatitis viruses. *Gastroenterol. Hepatol*. 5, 7-15

Waje, T., Dadah, A. J., Yusha'u, M., & Iliyasu, C. (2017). Demographic Study on Hepatitis A Infections among Outpatients of Selected Hospitals within Kaduna Metropolis, Nigeria. *International Archives of Public Health and Community Medicine*, *1*(1).

Wang, Z., Chen, Y., Xie, S., & Lv, H. (2016). Changing Epidemiological Characteristics of Hepatitis A in Zhejiang Province, China: Increased Susceptibility in Adults. *PLoS One*; 11(4):e0153804

World Health Organization. (2019). *Hepatitis virus A Fact Sheet*.